\documentclass[a4paper,reqno]{amsart}
\usepackage{amsfonts,epsfig}

\newtheorem{example}{Example}
\newtheorem{remark}{Remark}
\newtheorem{theorem}{Theorem}

\newcommand{\fr}[2]{\frac{\displaystyle{#1}}{\displaystyle{#2}}}

\newcommand{\cP}{{\mathcal P}}
\newcommand{\cL}{{\mathcal L}}

\newcommand{\cD}{{\mathcal D}}

\begin{document}
\title{
\bf Fractal Dimensions of the Hydrodynamic Modes of Diffusion}
\author{
T. Gilbert}
\address{Laboratoire de Physique Th\'eorique de la
Mati\`ere Condens\'ee, Universit\'e Paris VII, case~7020, Place
Jussieu, 75251 Paris Cedex 05, France}
\author{J. R. Dorfman}
\address{Department of Physics and Institute for Physical Science
and Technology, University of Maryland, College Park, MD 20742,
USA}
\author{P. Gaspard}
\address{Center for Nonlinear Phenomena and Complex Systems,
Universit\'e Libre de Bruxelles, Campus Plaine, C. P. 231, Blvd du
Triomphe, B-1050 Brussels, Belgium}
\date{\today}

\begin{abstract}
We consider the time-dependent statistical distributions of diffusive processes
in relaxation to a stationary state for simple, two dimensional chaotic
models based upon random walks on a line. We show that
the cumulative functions of the hydrodynamic modes of diffusion form fractal
curves in the complex plane, with a Hausdorff dimension larger than one.
In the limit of vanishing wavenumber, we derive a simple expression of the diffusion
coefficient in terms of this Hausdorff dimension and the positive Lyapunov
exponent of the chaotic model.
\end{abstract}
\maketitle

\section{Introduction}

The statistical mechanics of nonequilibrium processes has been
the subject of renewed attention. Computer studies have shown that the
classical motions of systems with large numbers of particles
are typically chaotic
\cite{dellago96,dorfman99,evans90,evans93,gaspard98,posch88,sinai79,
vanbeijeren95,vanbeijeren96,vanbeijeren97}.
These observations have motivated the theoretical study
of low-dimensional model systems where one can obtain quantitative
information about nonequilibrium processes in terms of the
chaotic properties of the microscopic dynamics of the systems.

In this context, several relationships have been established
between transport properties such as diffusion or viscosity and
characteristic quantities of chaos such as Lyapunov exponents, 
Kolmogorov-Sinai (KS) entropy per unit time, and fractal dimensions
\cite{baranyai93,chernov93a,chernov93,dorfman95,evans90b,gaspard90,gaspard95c}.
These relationships have been obtained for two different types of dynamical
systems: (a) deterministic systems with Gaussian thermostats and, (b) open 
Hamiltonian systems with absorbing boundaries.

In the deterministic systems with Gaussian thermostats \cite{evans90,hoover91},
the forces ruling the motion of the particles are modified 
in order to mimic a heat pump removing the excess energy introduced 
by the external fields or other nonequilibrium constraints.  
The modified dynamical system is still time-reversal symmetric but 
volumes are no longer preserved in phase space so that trajectories, 
in general, converge toward a {\em fractal attractor} which 
has an invariant Sinai-Ruelle-Bowen (SRB) measure,  absolutely
continuous with respect to Lebesgue measure in unstable directions and
fractal in stable directions.  On average, the phase-space volumes contract at 
a rate
which increases with the external fields and, thus, with the dissipation of
heat.  Since the overall contraction rate is equal to the sum of all the
Lyapunov exponents, the dissipation turns out to be related to the
characteristic quantities of chaos.  In this way, formulae have been obtained
for the viscosity of an externally sheared fluid
\cite{evans90b} and also for the diffusion of a Gaussian-thermostated Lorentz
gas \cite{baranyai93}.  In this last case, the formula relating the
diffusion coefficient to the sum of the Lyapunov exponents was also
rigorously proved \cite{chernov93a,chernov93}.
In the thermostated Lorentz gas, the kinetic energy is constant (instead of the
total energy) and the phase space of the flow has thus the dimension three.
The Lyapunov exponents of this system are $\lambda^+>0>\lambda^-$ with
$\vert\lambda^-\vert > \lambda^+$ for a nonvanishing field force $F$.
In the three-dimensional phase space, the fractal attractor ${\mathcal A}_F$
has a KS entropy equal to the positive Lyapunov exponent $h_{\rm
KS}=\lambda^+$ according to Pesin's equality, and its information dimension
is given by Young's formula: $D_{\rm I}=2+h_{\rm KS}/\vert\lambda^-\vert$.
In this expression for the fractal dimension, the 2 stands for the integer
dimensions of the flow direction and of the unstable direction which both are
equal to one for the attractor of a flow.  However, the attractor is fractal in
the stable direction which contributes its partial information dimension
$d_{\rm I}=h_{\rm KS}/\vert\lambda^-\vert$ to the total information
dimension of the attractor ${\mathcal A}_F$.  The diffusion
coefficient of the Gaussian-thermostated Lorentz gas at temperature $T$ has
been
alternatively expressed as
\begin{eqnarray}
{\mathcal D} &=& \lim_{F\to 0} \left(\frac{k_{\rm B}T}{F}\right)^2 \ \left[
-\lambda^-({\mathcal A}_F)-\lambda^+({\mathcal A}_F)\right] \nonumber\\ &=&
\lim_{F\to 0}
\left(\frac{k_{\rm B}T}{F}\right)^2 \ \left[ \vert\lambda^-({\mathcal
A}_F)\vert -h_{\rm KS}({\mathcal
A}_F)\right] \nonumber\\  &=& \lim_{F\to 0}
\left(\frac{k_{\rm B}T}{F}\right)^2 \  \lambda^+({\mathcal
A}_F)\ c_{\rm I}({\mathcal A}_F)\; ,
\label{eq:evans90b}
\end{eqnarray}
where the first form is a consequence of the average contraction of
phase-space volumes due to the Gaussian thermostatting mechanism and the 
relation between the rate of contraction and the rate
of entropy production in systems with Gaussian thermostats, the second
form follows from the first one by Pesin's equality, and the third one is
obtained by Young formula for the partial information codimension $c_{\rm
I}=1-d_{\rm I}=1-h_{\rm KS}/\vert\lambda^-\vert$ and from the fact that
$\lim_{F\to 0}\vert\lambda^-({\mathcal A}_F)\vert =\lim_{F\to 0}
\lambda^+({\mathcal A}_F)$ \cite{baranyai93,chernov93a,chernov93,dettmann00}.  In the
limit of arbitrarily small external field $F$, the information dimension of the
fractal attractor approaches the phase-space dimension in a way controlled by the
diffusion coefficient, which is the basis of this kind of formulae.

Very similar formulae have been obtained
for open Hamiltonian systems with absorbing
boundaries \cite{dorfman95,gaspard90,gaspard95c}.  Such boundaries are
naturally introduced in systems of scattering type in which particles or
trajectories undergo transient collisions before escaping out of a
phase-space region of varying geometry.  The absorbing boundaries have the
effect of driving the system out of equilibrium without a modification of
the forces which can remain Hamiltonian, i.e., time-reversal symmetric and
also volume preserving.  The escape of trajectories leads to the formation
of a {\em fractal repeller}, described by a invariant Gibbs measure which is fractal in both
the unstable and the stable directions.  For such systems, the rate of escape
is given as the difference between the sum of positive Lyapunov exponents and
the KS entropy \cite{eckmann85,kantz85}, calculated with respect to
the invariant Gibbs measure on the repeller.  This escape-rate formula was
rigorously proved for Anosov maps with rectangular holes
\cite{chernov97a,chernov97b}.  For the conservative Lorentz gas with absorbing
boundaries separated by a distance $L$, the Lyapunov exponents are
$\lambda^+ > 0 > \lambda^-=-\lambda^+$ and the diffusion coefficient has been
obtained by the following formulae
\cite{gaspard90,gaspard95c} where the quantities are evaluated for the fractal
repeller ${\mathcal R}_L$:
\begin{eqnarray}
{\mathcal D} &=& \lim_{L\to \infty} \left(\frac{L}{\pi}\right)^2 \ \left[
\lambda^+({\mathcal R}_L)-h_{\rm KS}({\mathcal R}_L)\right] \nonumber\\ &=&
\lim_{L\to \infty} \left(\frac{L}{\pi}\right)^2 \  \lambda^+({\mathcal
R}_L)\ c_{\rm I}({\mathcal R}_L) \nonumber\\ &=&
\lim_{L\to \infty} \left(\frac{L}{\pi}\right)^2 \  \lambda^+({\mathcal
R}_L)\ c_{\rm H}({\mathcal R}_L)\; ,
\label{eq:gaspard90}
\end{eqnarray}
where $c_{\rm I}=1-h_{\rm KS}/\lambda^+$ denotes the partial information
codimension in the stable or unstable directions, which are equivalent.  This
information codimension can be replaced by the partial Hausdorff codimension
$c_{\rm H}$ of the fractal repeller in the limit $L\to\infty$
\cite{gaspard95c}.  According to Eq. (\ref{eq:gaspard90}), the fractal
repeller fills the phase space in the limit where the
absorbing boundaries are sent to infinity.  The way the dimension of the
fractal repeller approaches the phase-space dimension is controlled by the
diffusion coefficient, and leads to the relationship between the transport
coefficient and the codimension.  The formula
(\ref{eq:gaspard90}) has been generalized to all the transport coefficients
\cite{dorfman95}.

The similarities between formulae (\ref{eq:evans90b}) and
(\ref{eq:gaspard90}) have been discussed for several years \cite{dettmann00}.  In particular, a
generalization of these formulae in the presence of both an external field and
absorbing boundaries was obtained \cite{tel96}.  However, each of the
two formulae depends very much on the type of invariant set, an
attractor or a repeller, and on the invariant measure, SRB or Gibbs, in
terms of which they are derived. Thus, since the corresponding
measures and fractal structures are so different in the two cases, it
has not yet been possible to provide satisfying reasons why the formulae are so
similar. 

The purpose of the present paper is to introduce a unifying approach which
transcends the aforementioned differences and in which a third and new formula
can be derived which expresses the diffusion coefficient in terms of the positive
Lyapunov exponent of the system and the Hausdorff dimension of a
fractal curve that describes a generalized hydrodynamic mode of the
diffusion process. This Hausdorff dimension is not, therefore, the dimension of an
attractor or a repeller.  The novelty of the present unifying approach is
that we consider an abstract fractal curve directly associated with the
hydrodynamic modes of relaxation toward a stationary state such as the equilibrium state.  
These hydrodynamic modes exist whether the system is thermostated or not, on the
condition that its spatial extension is large compared with
characteristic microscopic lengths, in order to sustain a process
of transport by diffusion.  We suppose that the time evolution of
the concentration of tracer particles can be decomposed by a spatial Fourier
transform into modes characterized by a wavenumber $k$ or, equivalently, a
wavelength $L=2\pi/k$.  Each mode describes the inhomogeneities of
concentration having this specified spatial periodicity and the time evolution
of a non-periodic concentration is obtained by superposition of all the modes.
Each mode has an exponential relaxation rate.  For a deterministic dynamical
system, these hydrodynamic modes can be defined in analogy with the
conditionally invariant measures \cite{pianigiani79} by compensating the
exponential decay of the mode amplitude with an appropriate renormalization at
each time step.  In the present context, the conditionally invariant
measures defining the hydrodynamic modes are complex because of the Fourier
transform.  A theory of these hydrodynamic modes has been described elsewhere
\cite{gaspard96,gaspard98}, where these conditionally invariant measures are
obtained as the eigendistributions of a Perron-Frobenius operator of the
dynamical system.  The main result we have to keep in mind here is that these
complex measures are singular with respect to the Lebesgue measure so that
they do not have density functions.  Instead, they have cumulative functions
defined by the measure of a variable set in phase space.  These cumulative
functions are complex, continuous and nondifferentiable.  When plotted in the
complex plane, they depict fractal curves of von Koch's
type \cite{edgar93,vonkoch04,vonkoch06} with Hausdorff dimensions between
one and two.  A recent work by the present authors \cite{gilbert00c} 
has showed that the singular character of the hydrodynamic modes 
accounts for the entropy production of irreversible thermodynamics.

In the present paper, we shall show that the Hausdorff dimension of the fractal
curves associated with the hydrodynamic modes is controlled by the diffusion
coefficient, which leads to a new relationship between transport and chaos.
We shall here prove our result for a whole class of
multi-baker maps.  This may appear restrictive but several important
results were first proved for the multi-baker before being extended to
more general chaotic systems.  In particular, let us mention that the
multi-baker models have been used for the study of the nonequilibrium
steady states and the entropy production
\cite{breymann96,breymann98,gaspard97,gilbert99b,gilbert99,vollmer97,vollmer98}.

The multi-baker models have the advantage of being exactly solvable so
that they are therefore appropriate for the intensive studies of
different nonequilibrium properties.  Many of their properties are known
in detail \cite{gilbert99d,tasaki98}.

The organization of the paper is as follows. In Sec.
\ref{s_models}, we discuss general models of deterministic random
walks, which include both conservative and so-called dissipative cases. In
Sec. \ref{s_statens}, the time evolution operator on the
statistical ensembles of trajectories is introduced. The
eigendistributions and their cumulative functions are derived.
Two different types of stationary solutions to the time evolution
operator are discussed in Sec. \ref{s_SS}, one for conservative
systems, the other for dissipative ones. The computation of the
Hausdorff dimensions of the cumulative functions of the
eigendistributions is done in Sec. \ref{s_HD}. Conclusions are
drawn in Sec. \ref{s_conc}.

\section{Deterministic Models of Diffusion \label{s_models}}

The simplest examples of reversible dynamical systems with
diffusive properties are based upon simple stochastic processes.
The most commonly studied is the discrete symmetric dyadic random walk
where independent particles are allowed to move on the sites of a
one-dimensional discrete lattice and the time evolution 
is determined by the condition that the probability is the same for a particle 
to move to the nearest neighboring site, on the right or left
\cite{gaspard92,gaspard97,gaspard98,tasaki94,tasaki95}.

In order to study this process as one with a phase-space dynamics, we
consider the set of all the possible infinite (both in the past
and future) trajectories of such particles. It consists of
sequences $\{n_i\}_{i=-\infty}^{+\infty}$ of integers $n_i\in
\mathbb Z$ with the restriction that the difference between the
integers at successive time steps is always equal to plus or minus
one, i.e. $\forall i \in \mathbb Z,$ $$n_i - n_{i - 1} = \pm
1.$$ The stochastic time evolution is thus replaced by the shift
operator on those sequences, $$\Sigma\big(\{n_i\}\big) =
\{n'_{i}\},$$ with $n'_i = n_{i+1}.$

Alternatively, a convenient representation of phase space is to
consider binary sequences of zeros and ones labeling respectively
hops to the left and right. A trajectory of the random walker is
then given by an integer coordinate representing, say, the
position at time zero and an arbitrary binary sequence
$\{\omega_i\}_{i = -\infty}^{+\infty},$ $\omega_i \in \{0,1\}
\:\forall i,$ coding all its (past and future) displacements.

This phase space is a two-dimensional continuum, which is best
represented by points on the unit square. The transposition from
symbolic sequences to points of the unit square is given by the
dyadic expansion of $x$ and $y.$ Let
\begin{eqnarray}
x &=& \sum_{i = 0}^\infty \frac{\omega_i}{2^{i +
1}},\label{x2exp}\\ y &=& \sum_{i = -1}^{-\infty}
\frac{\omega_i}{2^{-i}}\label{y2exp},
\end{eqnarray}
i.e. the point $(x,y)$ on the unit square is coded by the
bi-infinite sequence $\{\omega_i\}_{i = -\infty}^{+\infty},$ with
the positive indices (including zero) coding the $x$-component and
the negative ones the $y$-component.

The reason for introducing this representation of trajectories is
that the time evolution represented by the shift operator on
binary sequences is isomorphic to the baker-map acting on points
$(x,y)$ of the unit square,
\begin{equation}
(x,y)\rightarrow\left\{
\begin{array}{l@{\quad}l}
(2 x, y/2),&0\leq x < 1/2,\\ (2 x - 1, (y + 1)/2),&1/2\leq x<1.
\end{array}
\right. \label{bakermap}
\end{equation}
The proof that the shift on binary sequences and the baker map are
isomorphic is straightforward~: the baker map, Eq.
(\ref{bakermap}), takes $(x,y)$ to $(x',y')$ with the dyadic
expansion
\begin{eqnarray}
x' &=& \sum_{i = 0}^\infty \frac{\omega_{i+1}}{2^{i + 1}},\\ y'
&=& \sum_{i = -1}^{-\infty} \frac{\omega_{i+1}}{2^{-i}}.
\end{eqnarray}
In other words, the couple $(x',y')$ is coded by the shifted
sequence $\{\omega'_i\}_{i = -\infty}^{+\infty},$ with $\omega'_i
= \omega_{i + 1}.$

In order to describe the one-dimensional dyadic symmetric random walk,
we need to add the aforementioned integer $n$ which labels the site of
the one-dimensional lattice where the particle is located at the
current time.  Accordingly, this random walk is isomorphic to the {\em
symmetric dyadic multi-baker map} \cite{tasaki94,tasaki95}
\begin{equation}
B_2:\:(n,x,y)\rightarrow
\begin{cases}
(n - 1, 2 x, y/2),&0\leq x < 1/2,\\ (n + 1, 2 x - 1, (y +
1)/2),&1/2\leq x<1,
\end{cases}
\label{mbakermap}
\end{equation}
acting in the phase space ${\mathbb Z} \times [0,1]^2.$

Likewise, a one-dimensional non-sym\-metric $r$-adic random walk,
where a particle located at site $n$ has probability $p_j$ to hop
to the site $n + j$, with the integer $j$ taking values between
$-(r - 1)/2$ and $(r -1)/2$ (assuming $r$ to be odd\footnote{The
case $r$ even is treated similarly, with the modification that the
probability $p_0$ for a particle to remain at the same position is
set to zero.}) and $p_{- (r - 1)/2} + \ldots + p_{(r - 1)/2}=1$,
can also be mapped on a deterministic multi-baker map.

Similarly to the dyadic process, we code the trajectories of such
a process by sequences $\{\omega_i\}_{i = -\infty}^{+\infty},$
with $\omega_i\in\{0,\ldots,r - 1\}.$ Since every symbol
$\omega_i$ has respective weight $p_{\omega_i - (r - 1)/2},$ the
$r$-adic expansion of $x$ and $y$ takes a form slightly more
complicated than Eqs. (\ref{x2exp})-(\ref{y2exp}). We can write
those expansions under the recursive form
\begin{eqnarray}
x(\omega_0,\omega_1,\ldots) &=& \sum_{j < \omega_0} p_{[j - (r -
1)/2]} + p_{[\omega_0 - (r - 1)/2]}\;
x(\omega_1,\omega_2,\ldots),\label{xrexp}\\
y(\omega_{-1},\omega_{-2},\ldots) &=& \sum_{j < \omega_{-1}} p_{[(r
- 1)/2-j]} + p_{[(r - 1)/2-\omega_{-1}]}\;
y(\omega_{-2},\omega_{-3},\ldots).\label{yrexp}
\end{eqnarray}
The $r$-adic multi-baker map that mimics the random process is
given by
\begin{equation}
B_r:\:(n, x, y) \rightarrow \left\{
\begin{aligned}
\Big(n - \frac{r - 1}{2},& \frac{x}{p_{- (r - 1)/2}}, p_{(r -
1)/2}\; y\Big),\\ & 0\leq x < p_{- (r - 1)/2},\\ \Big(n - \frac{r -
3}{2},& \frac{x - p_{- (r - 1)/2}}{p_{- (r - 3)/2}}, p_{(r - 3)/2}
\; y + p_{(r - 1)/2}\Big),\\ &p_{- (r - 1)/2}\leq x < p_{- (r - 1)/2}
+ p_{- (r - 3)/2},\\ \vdots&\\ \bigg(n + \frac{r - 1}{2},& \frac{x
- \sum_{j = - (r - 1)/2}^{(r - 3)/2}p_j}{p_{(r - 1)/2}}, p_{-(r
- 1)/2} \; y + \sum_{j = -(r - 3)/2}^{(r - 1)/2}p_j \bigg),\\ & \sum_{j
= - (r - 1)/2}^{(r - 3)/2}p_j \leq x < 1.
\end{aligned}
\right. \label{radicbaker}
\end{equation}
The action of $B_r$ on points $x$ and $y$ with $r$-adic expansions
given by Eqs. (\ref{xrexp})-(\ref{yrexp}) is easily found to yield
\begin{multline}
B_r\big[ n,
x(\omega_0,\omega_1,\ldots),y(\omega_{-1},\omega_{-2},\ldots)\big]
=\\ \big[ n + \omega_0 - (r - 1)/2,
x(\omega_1,\omega_2,\ldots),y(\omega_{0},\omega_{-1},\ldots)\big].
\end{multline}

Notice the reversed order of the $p_i$'s along the $y$-coordinate
in Eq. (\ref{radicbaker}). This ensures the time
reversibility of the map : The involution $T(x,y) = (1 - y,1 - x)$
is a {\em time-reversal operator} or {\em reversal symmetry} for
$B_r$ in the sense that
\begin{equation}
T \circ B_r \circ T = B_r^{-1}. \label{treversal}
\end{equation}

The multi-baker map, Eq. (\ref{radicbaker}), is chaotic with the mean positive
Lyapunov exponent
\begin{equation}
\lambda^+ = \sum_{j= -(r - 1)/2}^{(r - 1)/2} p_j\; \ln\fr{1}{p_j}\; = \;
h_{\rm KS}
\; >\; 0 \; ,
\label{poslyap}
\end{equation}
which is equal to its Kolmogorov-Sinai entropy, and the mean negative
Lyapunov exponent
\begin{equation}
\lambda^- = \sum_{j= -(r - 1)/2}^{(r - 1)/2} p_j\; \ln p_{-j}
\; < \; 0 \; ,
\label{neglyap}
\end{equation}
both evaluated under the forward dynamics.

The $r$-adic multi-baker maps (\ref{radicbaker}) constitute simplified
models of the Poincar\'e-Birkhoff mappings ruling the collision dynamics of
a point-like particle elastically bouncing on hard disks fixed in the
plane \cite{gaspard98,tel00}.  In particular, the multi-baker models
share many of the chaotic properties of these Lorentz-type billiards.

\begin{example}
\label{ex2brw} The dyadic biased random walk where a particle
moves to the left with probability $p_{-1}\equiv q$ and to the
right with probability $p_1 = 1 - q$ is coded by binary sequences
which are mapped onto the unit square by the expansion
\begin{eqnarray}
x &=& \omega_0 q + \sum_{i = 1}^\infty \omega_i q \prod_{l = 0}^{i
- 1} p_{(2\omega_l - 1)},\label{x2bexp}\\ y &=& \omega_{-1} (1 - q)
+ \sum_{i = -2}^{-\infty} \omega_i(1 - q) \prod_{l = -1}^{i + 1}
p_{(1 - 2\omega_l)}.\label{y2bexp}
\end{eqnarray}

This transposition yields the so-called reversible dissipative
multi-baker map \cite{gilbert99d,gilbert99b,tasaki98}
\begin{equation}
B^{\rm d}_2:\:(n,x,y)\rightarrow
\begin{cases}
(n - 1, x/q, (1 - q)y),&0\leq x < q,\\(n + 1, (x -q)/(1 - q), 1 -
q + q y),&q \leq x<1.
\end{cases}
\label{brwbakermap}
\end{equation}
\end{example}

\begin{example}
\label{ex3rw} The triadic totally symmetric random walk, with
$p_{-1} = p_0 = p_1 \equiv 1/3,$ was discussed by the authors in
\cite{gilbert00c}. Similarly to Eqs. (\ref{x2exp})-(\ref{y2exp}),
infinite sequences of zeros, ones and twos are mapped onto the
unit square by the expansion
\begin{eqnarray}
x &=& \sum_{i = 0}^\infty \frac{\omega_i}{3^{i +
1}},\label{x3exp}\\ y &=& \sum_{i = -1}^{-\infty}
\frac{\omega_i}{3^{-i}}\label{y3exp}.
\end{eqnarray}
The triadic multi-baker map is simply
\begin{equation}
B_3:\:(n,x,y)\rightarrow
\begin{cases}
(n - 1, 3 x, y/3),&0\leq x < 1/3,\\ (n, 3 x - 1, (y +
1)/3),&1/3\leq x < 2/3,\\(n + 1, 3 x - 2, (y + 2)/3),&2/3\leq x<1.
\end{cases}
\label{3bakermap}
\end{equation}
\end{example}

\begin{example}
\label{ex3brw} The biased triadic random walk, where a particle
hops to the left with probability $p_{-1} \equiv s_L,$ to the
right with probability $p_{1} \equiv s_R,$ and remains at its
position with probability $p_0 = s_N \equiv 1 - s_{L} - s_{R},$ is
one of the processes studied by T\'el, Vollmer, and Breymann
\cite{breymann96,breymann98,vollmer97,vollmer98}. The
transposition of the infinite symbolic sequences to the points of
the unit square is similar to Eqs. (\ref{x2bexp})-(\ref{y2bexp}) and
yields the reversible dissipative triadic multi-baker map
\begin{equation}
B^{\rm d}_3:\:(n,x,y)\rightarrow
\begin{cases}
(n - 1, x/s_L, s_R y),&0\leq x < s_L,\\ (n, (x - s_L)/s_N, s_N y +
s_R),&s_L\leq x < s_L + s_N,\\(n + 1, (x - s_L - s_N)/s_R, s_L y +
s_N + s_R),&s_L + s_N \leq x<1.
\end{cases}
\label{3brwbakermap}
\end{equation}
The equivalence between the non-area-preserving map (\ref{3brwbakermap})
and an area-preserving multi-baker map with an extra variable of energy was
discussed in Ref. \cite{tasaki99}.
\end{example}

\section{Time Evolution of Statistical Ensembles \label{s_statens}}

In order to study the relaxation to stationarity of a large
ensemble of random walkers, we restrict the one-dimensional
lattice to a ring of $L$ sites and impose periodic boundary
conditions, that is we identify site $0$ with site $L.$ The phase space is
thus restricted to ${\cL}\times[0,1]^2$ with $\cL =
\{1,\dots,L\}$.

The time evolution of a density  $\rho_t(n,x,y)$ is determined by
the Perron-Frobenius operator~: $\rho_{t + 1}(n,x,y) = \cP
\rho_t(n,x,y) =
\rho_t\big[B_r^{-1}(n,x,y)\big]/J\big[B_r^{-1}(n,x,y)\big]$, where
we have introduced the Jacobian of the transformation,
$J\big[B_r^{-1}(n,x,y)\big] = p_{i}/p_{-i}$ if $\sum_{j > i}p_j
\leq y < \sum_{j \geq i} p_j.$ The average number of particles in
a cylinder $\Omega$ of the $n^{\rm th}$ cell is given by $\mu_t(\Omega) =
\iint_\Omega dx\,dy\, \rho_t(n,x,y)$, with time evolution $\mu_{t
+ 1}(\Omega)= \mu_t\big[B_r^{-1}(\Omega)\big]$.

To display the fractal forms underlying the relaxation to
stationarity, we consider the cumulative function defined as the
measure of a cylinder $\Omega = [0,1]\times[0,y]$ inside the
$n^{\rm th}$ site, i.~e.
\begin{equation}
g_t(n,y) = \int_0^1 dx\,\int_0^y dy'\,\rho_t(n,x,y').
\label{defcumfunc}
\end{equation}
We can assume, without loss of generality, that the
distribution function $\rho_t(n,x,y)$ is
uniform with respect to $x$,
since the $x$-direction is uniformly expanding.  Thus the distribution
will become uniform along the $x$ direction,
for long times, independently of the initial conditions.  Thus we
can write $\mu_t(n,[0,x]\times [0,y]) = x\; g_t(n,y)$ and find that the
time evolution of the cumulative function, $g_t(n,y)$, is given by
\begin{equation}
g_{t+1}(n, y) = \left\{
\begin{aligned} p_{-(r - 1)/2}\; g_t \Big(n + \frac{r -
1}{2},& \frac{y}{p_{(r - 1)/2}}\Big),\\ & 0\leq y < p_{(r - 1)/2},
\\
p_{-(r - 1)/2} \; g_t \Big(n + \frac{r - 1}{2},1\Big)
&\\ + p_{-(r-3)/2} \; g_t\Big(n + \frac{r - 3}{2},& \frac{y - p_{(r -
1)/2}}{p_{(r - 3)/2}}\Big),\\ &p_{(r - 1)/2}\leq y < p_{(r - 1)/2}
+ p_{(r - 3)/2},\\ \vdots&\\\sum_{j = -(r - 1)/2}^{(r - 3)/2} p_j
\; g_t(n - j,1) &\\ + p_{(r - 1)/2} \; g_t\bigg(n - \frac{r - 1}{2},&
\frac{y - \sum_{j = - (r - 3)/2}^{(r - 1)/2}p_j}{p_{-(r -
1)/2}}\bigg),\\ & \sum_{j = - (r - 3)/2}^{(r - 1)/2}p_j \leq y <
1.
\end{aligned}
\right. \label{tevcumfunc}
\end{equation}

In order to identify the hydrodynamic modes and their
eigenfunctions, we consider $y = 1$ in Eq. (\ref{tevcumfunc}).
This yields the matrix equation
\begin{equation}
g_{t + 1}(n,1) = \sum_{j = -(r-1)/2}^{(r - 1)/2} p_j \; g_t(n - j,
1). \label{gt1}
\end{equation}
The eigenmodes of this equation have the form $\psi_k(n) = \exp(i
k n)$ with eigenvalue
\begin{equation}
\chi_k = \sum_{j = -(r-1)/2}^{(r - 1)/2} p_j \exp(-i k j).
\label{chik}
\end{equation}
Moreover, the values of the wavenumber $k$ are restricted to $k =
2\pi m/L,$ with $m \in {\mathbb Z} \pmod L,$ by the periodic
boundary conditions.

According to the above considerations, we may suppose that 
the cumulative function $g_t(n,y)$ of the deterministic multi-baker
map may be expanded as ~:
\begin{equation}
g_t(n,y) = \sum_k \chi_k^t \; a_k \; \psi_k(n) \; F_k(y)\; ,
\label{expcumfunc}
\end{equation}
where $a_k$ are coefficients set by the initial conditions and
$F_k(y)$ is solution of the system
\begin{equation}
F_k(y) = \left\{
\begin{array}{ll}
\fr{p_{-(r - 1)/2}}{\chi_k} \exp\Big(i k\frac{r - 1}{2}\Big)
F_k\Big(\frac{y}{p_{(r - 1)/2}}\Big),\\ 
\hspace{1.5in} 0\leq y < p_{(r - 1)/2},
\\
\frac{p_{-(r - 1)/2}}{\chi_k}\exp\Big(i k\frac{r - 1}{2}\Big) \\
\qquad + \frac{p_{-(r-3)/2}}{\chi_k} \exp\Big(i k\frac{r - 3}{2}\Big)
F_k\Big(\frac{y - p_{(r - 1)/2}}{p_{(r - 3)/2}}\Big),\\
\hspace{1.5in} p_{(r - 1)/2}\leq y < p_{(r - 1)/2} + p_{(r -
3)/2},\\ \qquad\vdots\\ \\ \sum_{j = -(r - 1)/2}^{(r - 3)/2}
\frac{p_j}{\chi_k}\exp(-i k j) \\ \qquad + \frac{p_{(r - 1)/2}}{\chi_k}
\exp\Big(-i k\frac{r - 1}{2}\Big) F_k\bigg(\frac{y - \sum_{j = - (r -
3)/2}^{(r - 1)/2}p_j}{p_{-(r - 1)/2}}\bigg),\\ 
\hspace{1.5in} \sum_{j = - (r - 3)/2}^{(r - 1)/2}p_j\leq y < 1,
\end{array}
\right. \label{Fk}
\end{equation}
where we used the identity $F_k(1) = 1$.

\begin{remark}
As long as
\begin{equation}
p_j/\chi_k < 1,\;\forall j, \label{contract}
\end{equation}
Eq. (\ref{Fk}) is a contracting functional equation of de Rham
type \cite{edgar93,derham57} with a unique continuous solution
corresponding to the
cumulative function of the hydrodynamic eigendistribution of
wavenumber $k$ of the Perron-Frobenius operator.
\end{remark}

On large spatial scales, the transport process in the multi-baker chain can be
modeled by the advection-diffusion equation \cite{tel00}
\begin{equation}
\partial_t\, c = - v \; \partial_n \, c + {\mathcal D} \; \partial_n^2 \, c
\, ,
\label{advectdiff}
\end{equation}
where $c=c(n,t)=g_t(n,y=1)$ is the particle concentration given by the
cumulative function at $y=1$, $v$ is the average velocity of the particles
drifting in a biased random walk, and $\mathcal D$ is the diffusion
coefficient.  The result (\ref{advectdiff}) is a consequence of the
decomposition (\ref{expcumfunc}) of the cumulative function in terms of the
hydrodynamic modes $\psi_k(n)=\exp(ikn)$.   Indeed, if we substitute the
decomposition (\ref{expcumfunc}) into Eq. (\ref{advectdiff}), we find that the
hydrodynamic modes $c(n,t)=\chi_k^t \psi_k(n)$ are approximate solutions
under the approximation that the terms in $k^3$, $k^4$,... are
neglected, in which case we find that
\begin{equation}
\ln \chi_k = - i \; v \; k - {\mathcal D} \; k^2 + O(k^3) \; .
\label{dispersion}
\end{equation}
This approximation is justified for the advection-diffusion equation
(\ref{advectdiff}) because the large-scale limit is equivalent to the small
wavenumber limit $k\to 0$.  In this way, an expansion of the eigenvalue 
(\ref{chik}) in powers of the wavenumber $k$ allows us to identify the drift velocity:
\begin{equation}
v \equiv \sum_{j = -(r - 1)/2}^{(r - 1)/2}j\; p_j\; ,\label{drift}
\end{equation}
and the diffusion coefficient:
\begin{equation}
{\mathcal D} \equiv \fr{1}{2}\sum_{j = -(r - 1)/2}^{(r - 1)/2} j^2\; p_j -
\fr{1}{2}\left(\sum_{j = -(r - 1)/2}^{(r - 1)/2} j \; p_j\right)^2 \; .
\label{difcoeff}
\end{equation}
in terms of the parameters of the mapping (\ref{radicbaker}).

\begin{remark}
For a symmetric random walk, with $r$ odd, and where $p_j = p_{-j}\;\forall j,$ the
eigenvalue is real and Eq. (\ref{chik}) becomes~:
\begin{equation}
\chi_k = p_0 + 2 \sum_{j = 1}^{(r - 1)/2} p_j \; \cos(k j).
\label{chiksymcase}
\end{equation}
In this symmetric case, the drift velocity vanishes, $v=0$, and the random
walk is
purely diffusive with the diffusion coefficient
\begin{equation}
{\mathcal D} = \sum_{j = 1}^{(r - 1)/2} j^2 \; p_j \; .
\label{symdiffcoeff}
\end{equation}
For totally symmetric $r$-adic random walks for which $p_j=1/r$ $\forall j$,
the eigenvalue is given by
\begin{equation}
\chi_k = \frac{\sin(kr/2)}{r\sin(k/2)},
\label{chiksymcaseradic}
\end{equation}
and the diffusion coefficient is
\begin{equation}
{\mathcal D} = \frac{r^2-1}{24}\; .
\label{symdiffcoeffradic}
\end{equation}
\end{remark}

For small $k$, we can expand $F_k$ in powers of $k$ and obtain the
equivalent of a gradient expansion~:
\begin{equation}
F_k(y) = F_0(y) + i k T(y) + O\big(k^2\big). \label{Fkexp}
\end{equation}
The first term of this equation is a real function $F_0$ and corresponds
to the stationary state of eigenvalue 1. In the next section, we
will differentiate between regular and singular stationary
solutions. The second term in Eq. (\ref{Fkexp}) is purely
imaginary, i.e. $T$ is a real function. Note that
\begin{equation}
\fr{1}{\chi_k} =  1 + i k v + O\big(k^2\big),
\label{expchik}
\end{equation}
where $v$ is the drift velocity (\ref{drift}) of the random walker
due to a possible bias. Substituting Eq. (\ref{Fkexp}) into Eq.
(\ref{Fk}) and, using Eq. (\ref{expchik}), we find the general
expression satisfied by $T(y)$~:
\begin{equation}
\label{genTakagi} T(y) = \left\{
\begin{array}{ll}
p_{-(r - 1)/2}\Bigg[\Big(\frac{r - 1}{2} + v\Big)
F_0\Big(\frac{y}{p_{(r - 1)/2}}\Big) + T \Big(\frac{y}{p_{(r -
1)/2}}\Big)\Bigg],\\ &\hspace{-.5cm} 0\leq y < p_{(r - 1)/2},
\\
p_{-(r - 1)/2}\Big(\frac{r - 1}{2} + v\Big)
\\ \qquad + p_{-(r - 3)/2}\Bigg[\Big(\frac{r - 3}{2} + v\Big)
F_0\Big(\frac{y - p_{(r - 1)/2}}{p_{(r - 3)/2}}\Big) \\ \qquad
\qquad + T \Big(\frac{y - p_{(r - 1)/2}}{p_{(r -
3)/2}}\Big)\Bigg],\\ &\hspace{-3.1cm}p_{(r - 1)/2}\leq y < p_{(r -
1)/2} + p_{(r - 3)/2},\\ \qquad\vdots\\ \\ \sum_{j = -(r -
1)/2}^{(r - 3)/2} p_j\Big( -j + v\Big) \\ \qquad + p_{(r -
1)/2}\Bigg[ \Big(-\fr{r - 1}{2} + v\Big)F_0\bigg(\frac{y -
\sum_{j = - (r - 3)/2}^{(r - 1)/2}p_j}{p_{-(r - 1)/2}}\bigg) \\
\qquad\qquad + T\bigg(\frac{y - \sum_{j = - (r - 3)/2}^{(r -
1)/2}p_j}{p_{-(r - 1)/2}}\bigg)\Bigg],\\ &\hspace{-1.6cm} \sum_{j
= - (r - 3)/2}^{(r - 1)/2}p_j \leq y < 1.
\end{array}
\right.
\end{equation}

In Sec. \ref{s_SS}, we will make the connection between this
equation and the functional equation defining a function introduced by Takagi
\cite{takagi03}.

\section{Stationary State \label{s_SS}}

The stationary eigenstate of the Perron-Frobenius operator has
eigenvalue 1, with a cumulative function $F_0$ given by Eq. (\ref{Fk}) for a
vanishing wavenumber $k=0$~:
\begin{equation}
\label{F0} F_0(y) = \left\{
\begin{array}{ll}
p_{-(r - 1)/2} \; F_0\Big(\fr{y}{p_{(r - 1)/2}}\Big),\qquad&0\leq y <
p_{(r - 1)/2},\\ p_{-(r - 1)/2} + p_{-(r - 3)/2} \; F_0\Big(\fr{y -
p_{(r - 1)/2}}{p_{(r - 3)/2}}\Big),\\&\hspace{-2cm}p_{(r -
1)/2}\leq y < p_{(r - 1)/2}+ p_{(r - 3)/2},\\ \text{etc.}
\end{array}
\right.
\end{equation}

For a symmetric process, where $p_j = p_{-j}\;\forall j,$ the
solution to this equation is simply $F_0(y) = y,$ corresponding to
the uniform density of the equilibrium stationary state.

\begin{example}
The dyadic symmetric random walk has the stationary state given by $F_0(y) =
y$. Equation (\ref{genTakagi}) thus becomes
\begin{equation}
T(y) =
\begin{cases}
y + \frac{1}{2} T(2 y),&0\leq y < \frac{1}{2},\\  1 - y + \frac{1}{2} T(2 y
-1),
&\frac{1}{2} \leq y < 1.
\end{cases}
\label{2rwtakagi}
\end{equation}
The solution to this functional equation is the Takagi function
\cite{takagi03,tasaki95}, which justifies the name generalized
Takagi function for $T$ in Eq. (\ref{genTakagi}). The Takagi function is
known to
be a nowhere differentiable function with Hausdorff dimension
$D_{\rm H}(T) = 1$.  As will be seen later, this property is shared by
the whole class of functions $T$ given by Eq. (\ref{genTakagi}).
\end{example}

For a non-symmetric process, the contraction and expansion factors
differ so that, in this case, $F_0$ is a Lebesgue singular function
\cite{tasaki98}.

\begin{figure}
\centerline{\epsfig{file=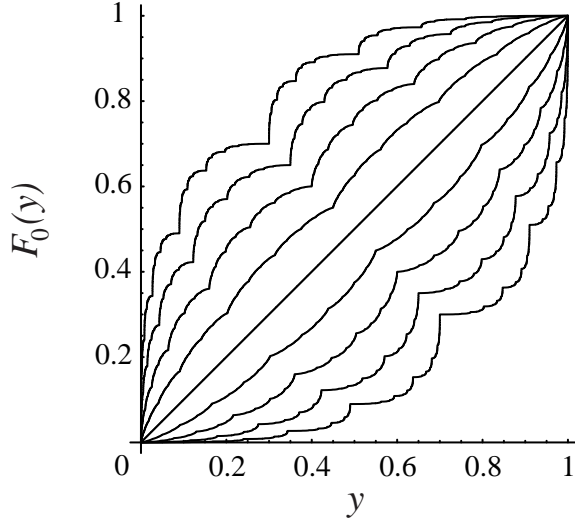,height=70mm}}
\caption{
For the biased random walk of Example \ref{ex2brw}, $F_0$ is
a Lebesgue singular function, obtained as the solution of Eq.
(\ref{brwF0}), and which we display for values of $q$ ranging 
from $0.3$ (bottom curve) to $0.7$ (top curve). 
The straight line corresponds to $q=1/2$. Each curve was computed at
$2^{10}$ points connected together by lines.\label{figbrwF0}}
\end{figure}

\begin{figure}
\centerline{\epsfig{file=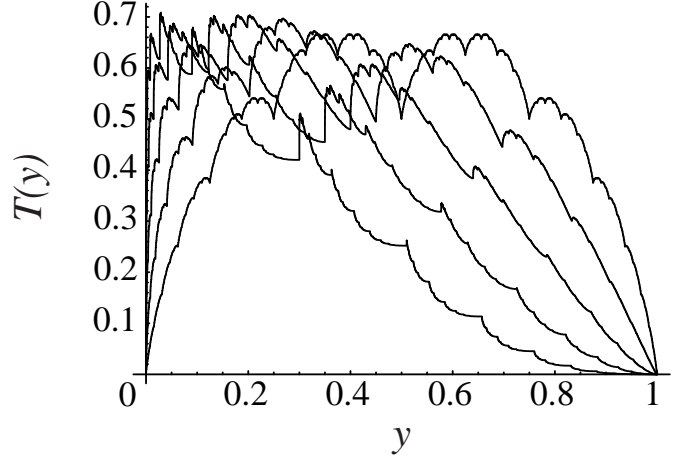,height=65mm}}
\caption{The generalized Takagi function for the biased random
walk, Eq. (\ref{brwT}). The parameter values range from $q= 0.5$
which gives the original Takagi function, Eq. (\ref{2rwtakagi}) (top curve
at the right) to $q = 0.7$ (bottom right).  Each curve was computed at
$2^{10}$ points connected together by lines. \label{figbrwT}}
\end{figure}

\begin{example}
The cumulative function of the stationary state of the reversible
dissipative multi-baker map, Eq. (\ref{brwbakermap}), of Example
\ref{ex2brw} satisfies
\begin{equation}
\label{brwF0} F_0(y) =
\begin{cases}
q F_0\Big(\frac{y}{1 - q}\Big),&0\leq y < 1 - q,\\ q + (1 - q)
F_0\Big(\frac{y - 1 + q}{q}\Big),&1 - q\leq y < 1.
\end{cases}
\end{equation}
We show in Fig. \ref{figbrwF0} this function evaluated for $q =
0.3, 0.35, \ldots, 0.7.$
For this map, the generalized Takagi function, Eq. (\ref{genTakagi}), is
found to
be a solution of
\begin{equation}
\label{brwT} T(y) =
\begin{cases}
2(1 - q) F_0(y) + qT\Big(\frac{y}{1 - q}\Big),&0\leq y < 1 - q,\\ 2
q[1 - F_0(y)] + (1 - q) T\Big(\frac{y - 1 + q}{q}\Big),&1 - q\leq
y < 1.
\end{cases}
\end{equation}
This function is depicted in Fig. \ref{figbrwT} for $q = 0.5$ to
$0.7.$ For values of $q$ lower than $0.5,$ we note that $T$ has
the symmetry $T[q](y) = T[1 - q](1 - y).$
\end{example}

\section{Hausdorff Dimension of the Hydrodynamic
modes\label{s_HD}}

In this section, we investigate the fractal properties of the
cumulative functions $F_k$ of the hydrodynamic modes for
$k\neq0$, as given by Eq. (\ref{Fk}), and we compute their Hausdorff
dimensions.

In order to compute the solutions of Eq. (\ref{Fk}),
we return to the symbolic dynamics and replace $y$ by the $r$-adic
expansion, Eq. (\ref{yrexp}). We consider points $y$ with finite
$r$-adic expansion $y = y(\omega_1,\ldots,\omega_l)$ and rewrite
Eq. (\ref{Fk}) in terms of the symbolic sequences~:
\begin{multline}
F_k[y(\omega_1,\ldots,\omega_l)] = \sum_{j < \omega_1} \frac{p_{[j
- (r - 1)/2]}}{\chi_k} \; \exp\{-i k [j - (r - 1)/2]\} \\
+ \frac{p_{[\omega_1 - (r - 1)/2]}}{\chi_k} \; \exp\{-i k [\omega_1 - (r
- 1)/2]\} \; F_k[y(\omega_2,\ldots,\omega_l)]. \label{Fksymbexp}
\end{multline}
Starting with $F_k[\emptyset]=0$, we can successively compute
$$F_k[y(\omega_1)],\quad F_k[y(\omega_1,\omega_2)],\quad
F_k[y(\omega_1,\omega_2,\omega_3)],\quad \text{etc}.$$
All these points are plotted in the complex plane where they form a fractal
curve $({\rm Re}F_k,{\rm Im}F_k)$ similar to von Koch's curve, under
certain conditions
(see the following examples).

\begin{example}
Consider the triadic totally symmetric random walk of Example
\ref{ex3rw}. From Eq. (\ref{chiksymcase}), the eigenvalues are found to
be $\chi_k = (1/3) (1 + 2 \cos k)$. The condition, Eq.
(\ref{contract}), that Eq. (\ref{Fk}) be contracting becomes $(1 +
2 \cos k)>1$, which is satisfied as long as $-\pi/2<k<\pi/2$. In
Fig. \ref{fig3rwFk}, we plot the imaginary versus real parts of
$F_k$ for $k = 0.1$ and $0.5$. The first of these two graphs for the smallest
wavenumber $k=0.1$, Fig.\ref{fig3rwFk}a, is dominated by the linear
contributions
to $F_k$, i.e. $F_k(y) \simeq y + i k T(y)$.  As we argued in
\cite{gilbert00c}, the first order term in $k$, $T$, is the
triadic equivalent of the Takagi function Eq. (\ref{2rwtakagi}).
In this case, $T(y)$ is solution of the functional equation
\begin{equation}
T(y) =
\begin{cases}
y + \frac{1}{3} T(3 y),&0\leq y < \frac{1}{3},\\ \frac{1}{3} + \frac{1}{3}
T(3 y -
1),&\frac{1}{3}\leq y < \frac{2}{3},\\ 1 - y + \frac{1}{3} T(3 y -
2),& \frac{2}{3}\leq y < 1.
\end{cases}
\label{3rwtakagi}
\end{equation}

The deviations -- observed in Fig.\ref{fig3rwFk}b -- of the curve $({\rm
Re}F_k,{\rm Im}F_k)$ with respect to the generalized Takagi function
(\ref{3rwtakagi}) are due to contributions which are second order in
the wavenumber $k$.  As we will show, these deviations are responsible for
the fact that the Hausdorff dimension of this curve is not one for non-zero
wavenumbers $k$. In fact, $D_{\rm H}(F_k)>1$.

\begin{figure}
\centerline{\epsfig{file=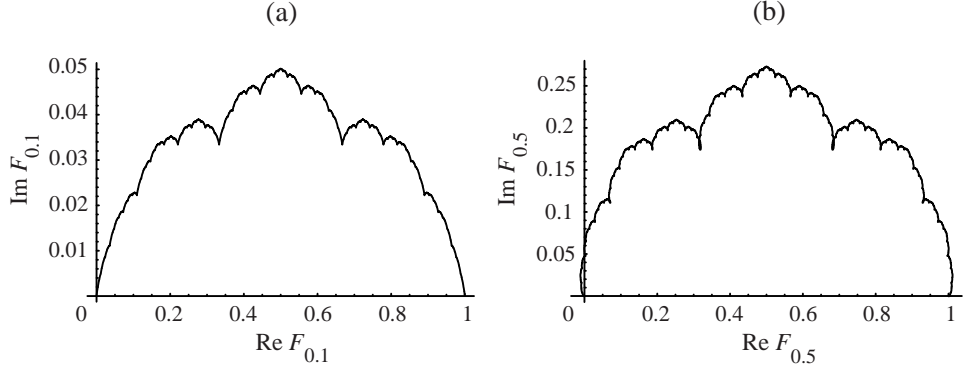,height=50mm}}
\caption{$F_k$ for the triadic totally symmetric random walk:
(a) $k=0.1$; (b) $k=0.5$.
The curves are computed at $3^{10}$ points connected together by lines.}
\label{fig3rwFk}
\end{figure}
\end{example}

\begin{example}
Still considering the triadic totally symmetric random walk, we
observe a rather spectacular fractal for the large wavenumber $k = \pi/3$
which corresponds to the eigenvalue $\chi_{\pi/3} = 2/3$. This case is
shown in Fig. \ref{fig3rwFpi3}. The points 0, 1/3, 2/3 and 1 are at the
vertices of a half hexagon, which is responsible for the apparent
hexagonal symmetry of this figure.
\begin{figure}[htb]
\centerline{\epsfig{file=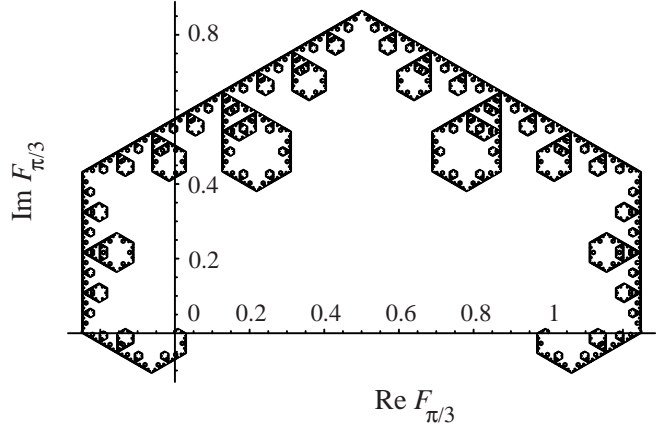,height=60mm}}
\caption{$F_{\pi/3}$ for the triadic totally symmetric random
walk. The curve is computed at $3^{10}$ points, and connected together
by lines. \label{fig3rwFpi3}}
\end{figure}
\end{example}

\begin{example}
Another interesting case corresponds to the critical value
$k_c$ of $k$ for which Eq. (\ref{Fk}) looses its contracting
property, i.e. $\chi_{k_c} = p_j$ in one of the conditions
(\ref{contract}). This
leads to interesting solutions of Eq. (\ref{Fk}). For instance, in
the triadic totally symmetric random walk, $k_c = \pi/2$ yields
\begin{equation}
F_{\pi/2}(y) = \begin{cases} -i F_{\pi/2}(3 y),&0\leq y < 1/3,\\
-i + F_{\pi/2}(3 y - 1),&1/3\leq y < 2/3,\\ 1 - i + i F_{\pi/2}(3
y - 2),&2/3\leq y <1.
\end{cases}
\end{equation}
$F_{\pi/2}(y)$ is thus a non-continuous function of $y$.  In the complex
plane, the curve disintegrates into a disjoint set of points forming the
square lattice ${\mathbb Z}^2$.  Figure \ref{fig3rwFpi2} displays this
set constructed with the $3^{10}$ points $y(\omega_1,\ldots,\omega_{10})$.
\begin{figure}[htb]
\centerline{\epsfig{file=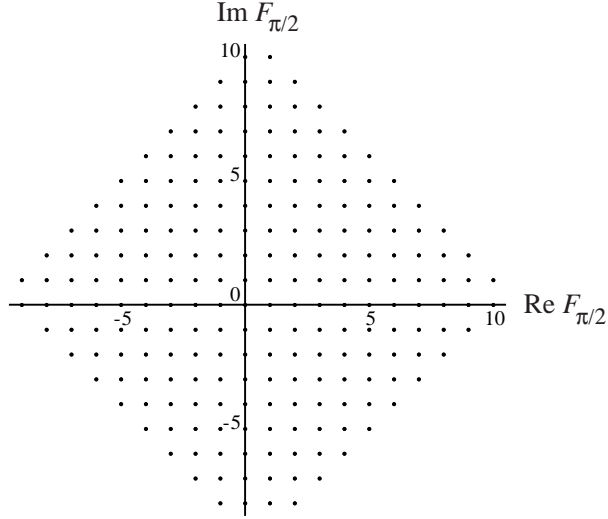,height=75mm}}
\caption{$F_{\pi/2}$ for the triadic totally symmetric random
walk. The set is composed of the points of ${\mathbb Z}^2.$
\label{fig3rwFpi2}}
\end{figure}
Similar functions can be observed for the 2-adic and 5-adic
totally symmetric random walks for $k_c = \pi/3$. In the 2-adic
case, the set is a regular triangular lattice; in the 5-adic case, a
regular hexagonal lattice.
\end{example}

We turn to the computation of the Hausdorff dimension of $F_k$, i.e., of
the curve $({\rm Re}F_k,{\rm Im}F_k)$ drawn in the complex plane.  The above
examples show that the plot of $({\rm Re}F_k,{\rm Im}F_k)$ forms a continuous
curve in the complex plane as long as $\vert k \vert <k_c$.  However, the
curve densely fills plain domains of the complex plane already for lower
values of the wavenumber, $k_f < \vert k \vert <k_c$.  For these values,
the Hausdorff dimension of the curve saturates at the dimension two of the
complex plane so that $D_{\rm H}(k)=2$ for $k_f \leq \vert k \vert <k_c$.
At the limiting value $\vert k \vert = k_f$, the plot of $({\rm Re}F_k,{\rm Im}F_k)$
forms a fractal curve similar to the L\'evy dragon which has a Hausdorff
dimension equal to two \cite{edgar93,levy38}.

For still lower values of the wavenumber $\vert k \vert < k_f$, we consider
the {\em Hausdorff measure}, $\Gamma_l^d(F_k),$ of the cylinder sets specified
by the sequences $\{\omega_1,\ldots,\omega_l\}$~:
\begin{equation}
\label{Hausmeas} \Gamma_l^d(F_k) = \sum_{\omega_1,\ldots,\omega_l}
\Big| \Delta F_k[y(\omega_1,\ldots,\omega_l)] \Big|^d,
\end{equation}
where $$\Delta F_k[y(\omega_1,\ldots,\omega_l)] =
F_k[y(\omega_1,\ldots,\omega_l + 1)] -
F_k[y(\omega_1,\ldots,\omega_l)],$$ with the notation
\begin{equation}
\{\omega_1,\ldots,\omega_l+1\} =
\begin{cases}
\{\omega_1,\ldots,\omega_{l - 1},\omega_l+1\},& \omega_l < r -
1,\\ \{\omega_1,\ldots,\omega_{l - 1}+1,0\},&\omega_l = r - 1,
\end{cases}
\end{equation}
and we make the further convention that $y(r-1,\ldots,r-1,r) = 1$.
The quantities $\Delta F_k[y(\omega_1,\ldots,\omega_l)]$ are the
segments of the curve, $({\rm Re}F_k,{\rm Im}F_k)$ at the $l^{\rm th}$
step of its construction. In the complex plane, the $l^{\rm th}$
approximant of the curve is covered by disks of diameter given by the
absolute value of the segments $\Delta
F_k[y(\omega_1,\ldots,\omega_l)]$, appearing on the right hand side of
Eq. (\ref{Hausmeas}).
We recall that the {\em Hausdorff dimension}, $D_{\rm H}(k),$ of $F_k$
is the value of $d$ such that \cite{eckmann85}
\begin{equation}
\label{Hausdim}
\lim_{l\rightarrow\infty} \Gamma_l^d(F_k) =
\begin{cases}
\infty,&d< D_{\rm H}(k),\\ 0,&d > D_{\rm H}(k). \end{cases}
\end{equation}

With the help of Eq. (\ref{Fksymbexp}), it is straightforward to
check that
\begin{eqnarray}
\Big|\Delta F_k[y(\omega_1,\ldots,\omega_l)] \Big|&=&
\fr{p_{[\omega_1-(r - 1)/2]}}{\vert\chi_k\vert}\Big|\Delta
F_k[y(\omega_2,\ldots,\omega_l)] \Big|,\nonumber\\ &=&
\fr{1}{{\vert\chi_k\vert}^l}\prod_{i = 1}^l p_{[\omega_i-(r -
1)/2]}.\label{DeltaFk}
\end{eqnarray}
Hence, the Hausdorff measure
\begin{equation}
\Gamma_l^d(F_k) = \fr{1}{{\vert\chi_k\vert}^{l d}}
\sum_{\omega_1,\ldots,\omega_l}
\prod_{i=1}^l
{p_{[\omega_i-(r - 1)/2]}}^{d} = \left( \fr{1}{{\vert\chi_k\vert}^{d}}
\sum_{j=-(r-1)/2}^{(r-1)/2} {p_j}^{d}\right)^l . \label{HausmeasFk}
\end{equation}
According to Eq. (\ref{Hausdim}), the Hausdorff dimension thus satisfies
\begin{equation}
\label{HausdimFk} \vert\chi_k\vert^{D_{\rm H}(k)} = p_{-(r - 1)/2}^{D_{\rm
H}(k)} + \dots + p_{(r - 1)/2}^{D_{\rm H}(k)}.
\end{equation}
\begin{example}
For the triadic totally symmetric random walk, the Hausdorff
dimension of $F_{\pi/3},$ shown in Fig. \ref{fig3rwFpi3},
satisfies $$\bigg(\fr{2}{3}\bigg)^{D_{\rm H}(\pi/3)} =
3\bigg(\fr{1}{3}\bigg)^{D_{\rm H}(\pi/3)}.$$ Therefore $D_{\rm H}(\pi/3) =
\ln(3)/\ln(2)$.
\end{example}

\begin{remark}
It is not possible to solve Eq. (\ref{HausdimFk}) and find a
general expression for $D_{\rm H}(k)$ for all processes.  However for the
case of totally symmetric $r$-adic random walks for which $p_j =
1/r,\;\forall j$, we can solve Eq. (\ref{HausdimFk}) and find
\begin{equation}
D_{\rm H}(k)=\begin{cases}
\fr{\ln(r)}{\ln(r \chi_k)} \; , & \vert k \vert < k_f \;
,\\ 2 \; , & k_f \leq \vert k \vert < k_c \; .\end{cases}
\label{symrwHausdim}
\end{equation}
$k_f$ is the wavenumber such that $\chi_k=1/\sqrt{r}$ where the curve starts
to densely fill the complex plane.  $k_c$ is the wavenumber such that
$\chi_k=1/r$, above which the cumulative function $F_k$ no longer
forms a continuous curve in the complex plane.  For the symmetric dyadic case
$r=2$, the Hausdorff dimension (\ref{symrwHausdim}) has been previously
derived \cite{tasaki93}.
\end{remark}

\begin{example}
A numerical computation of Eq. (\ref{symrwHausdim}) for the
totally symmetric triadic random walk is shown in Fig.
\ref{fig3rwHausdim}.  In this case, $D_{\rm H}(k)=\ln(3)/\ln(1+2\cos k)$ for
$\vert k \vert < k_f=1.1960...$, while $D_{\rm H}(k)=2$ for $k_f\leq \vert k
\vert < k_c=\pi/2$.
\begin{figure}
\centerline{\epsfig{file=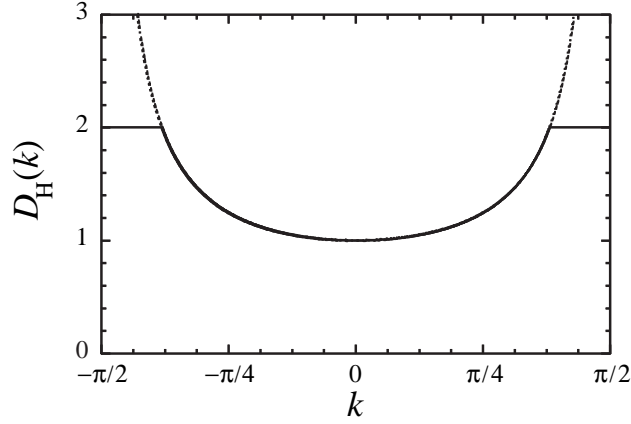,height=60mm}}
\caption{The Hausdorff dimension $D_{\rm H}$ for the
triadic totally symmetric random walk,
given by Eq. (\ref{symrwHausdim}) with $r=3$.
The dashed line depicts the extrapolation of $D_{\rm H}(k)$ outside of the
interval $\vert k \vert < k_f=1.1960...$.
\label{fig3rwHausdim}}
\end{figure}
\end{example}

For small wavenumbers, i.e. $|k| \ll k_f$, we can expand the absolute value of
the eigenvalue $\chi_k$, given by Eq. (\ref{chik}), in terms of the diffusion
coefficient (\ref{difcoeff}) of the general $r$-adic random walk~:
\begin{equation}
\vert\chi_k\vert = 1 - \cD k^2 + O(k^4). \label{chikexp}
\end{equation}
Notice that the absolute value of the eigenvalue does not depend on the drift
velocity $v$ in the linear and quadratic terms of its expansion in powers
of the wavenumber $k$.

To establish a connection between the diffusion coefficient and
the Hausdorff dimension of the hydrodynamic modes (for which
$|k|\ll k_f$), we let
\begin{equation}
D_{\rm H}(k) = 1 + a k^2 + O(k^4),
\label{deltaH}
\end{equation} and expand
\begin{equation}\label{expandright}
\sum_j p_j^{D_{\rm H}(k)} = 1 + a \; k^2 \; \sum_j p_j\; \ln p_j + O(k^4).
\end{equation}
Substituting Eqs. (\ref{chikexp}) and (\ref{expandright}) into Eq.
(\ref{HausdimFk}), we find
\begin{equation}
a = \fr{\cD}{\sum_j p_j\; \ln p_j^{-1}}= \fr{\cD}{\lambda^+}, \label{soldeltaH}
\end{equation}
where $\lambda^+$ is the positive Lyapunov exponent (\ref{poslyap}) of our
system.

We can thus state the following:
\begin{theorem}\label{theorem1}
For an $r$-adic multi-baker map with probabilities $\lbrace p_{-(r
- 1)/2}, \linebreak\ldots, p_{(r - 1)/2}\rbrace$, the cumulative
function of the hydrodynamic mode of wavenumber $k$, Eq.
(\ref{Fk}), forms in the complex plane a curve with the Hausdorff dimension
\begin{equation}
D_{\rm H}(k) = 1 + \fr{\cD}{\lambda^+} \; k^2 + O\big(k^4\big),
\label{Hausdimexp}
\end{equation}
where $\lambda^+$ is the positive Lyapunov exponent (\ref{poslyap}) of this
system
and $\cD$ is its diffusion coefficient as specified by Eq. (\ref{difcoeff}).
\end{theorem}

As a corollary, we have the following:
\begin{theorem}\label{theorem2}
Under the same conditions as in Theorem \ref{theorem1}, the diffusion
coefficient of
the system is given in terms of the positive Lyapunov exponent $\lambda^+$
and the
Hausdorff dimension $D_{\rm H}(k)$ of the cumulative function of the
hydrodynamic
mode of wavenumber $k$ according to
\begin{equation}
\cD = \lambda^+ \lim_{k\to 0} \frac{D_{\rm H}(k)-1}{k^2} . \label{formula}
\end{equation}
\end{theorem}

\begin{remark}
The expression (\ref{formula}) bears a relation to the formula
(\ref{eq:gaspard90})
giving the diffusion coefficient in terms of the Lyapunov exponent and the
Hausdorff codimension of the fractal repeller ${\mathcal R}_L$ of orbits
trapped in a 2-dimensional ordered Lorentz gas of size $L$ \cite{gaspard95c}.
The analogy is best seen by writing Eq. (\ref{formula}) for a hydrodynamic mode
of the smallest possible nonvanishing wavenumber $k=2\pi/L$ in a periodic
chain of length $L$, whereupon Eq.(\ref{formula}) takes the form
\begin{equation}
\cD = \lim_{L\rightarrow\infty}\Big(\fr{L}{2\pi}\Big)^2\lambda^+
\left[D_{\rm H}(2\pi/L)-1\right], \label{difcoefexp}
\end{equation}
where $D_{\rm H}(2\pi/L)$ is the dimension of the fractal curve $F_k$
with $k=2\pi/L$.  Although $a k^2$ in Eq. (\ref{deltaH}) plays a role very
similar to the codimension $c_{\rm H}({\mathcal R}_L)$ of Eq.
(\ref{eq:gaspard90}), it cannot be identified as such because it is the
coefficient of a correction to $D_{\rm H}$ {\em above} 1, which stems from
the definition of $F_k$ as a cumulative function.
\end{remark}

\section{Conclusions \label{s_conc}}

In this paper, we have shown with the help of simple models of
deterministic diffusion that, in the last stages of the relaxation toward the
stationary state, the hydrodynamic modes have cumulative functions with fractal
Hausdorff dimensions. For small enough wavenumbers, we were able to derive an
expression for the Hausdorff dimensions in terms of the diffusion
coefficient and positive Lyapunov exponent of the system.  Accordingly, the
diffusion coefficient can be computed in terms of the Hausdorff dimension and
the positive Lyapunov exponent, which is the content of Theorem
\ref{theorem2}.

Our formula (\ref{formula}) establishes a new relationship between the
diffusion coefficient and the characteristic quantities of chaos, which
generalizes both the equation (\ref{eq:evans90b}) from the thermostated-system
approach and the equation (\ref{eq:gaspard90}) from the escape-rate
formalism, as anticipated in the Introduction.  On the one hand, Eq.
(\ref{formula}) remains applicable whether the system is thermostated
(i.e., in the non-area-preserving case) or not (i.e., in the area-preserving
case).  On the other hand, no absorbing boundary is required in the
present approach where Eq. (\ref{formula}) was derived.  Accordingly, the
number of particles is kept constant in systems to which Eq. (\ref{formula})
applies.  In this regard, our new formula (\ref{formula}) unifies the
previous formulae (\ref{eq:evans90b}) and (\ref{eq:gaspard90}).
The novelty of the present approach is to consider the Hausdorff
dimension of an abstract curve in the complex plane, instead of the Hausdorff
(or information) dimension of either an attractor or a repeller.  This
abstract curve is the cumulative function of the hydrodynamic mode of
diffusion, which is a general concept of nonequilibrium statistical mechanics.

In view of the generality of the considerations developed in the present
paper, we conjecture that Eq. (\ref{Hausdimexp}) for the Hausdorff dimension
of the hydrodynamic modes extends to spatially extended Axiom-A
systems with two degrees of freedom which sustain transport by diffusion, and
more generally to such chaotic systems with two degrees of freedom.

Moreover, the similarity between Eq. (\ref{eq:gaspard90}) and Eq.
(\ref{difcoefexp}) of the escape-rate formalism suggests a generalization of
our formalism to other transport processes. As shown by Dorfman and Gaspard
\cite{dorfman95,gaspard95}, the escape-rate formalism generalizes to other
transport phenomena by setting up a first
passage problem in the space of the Helfand moment
\cite{helfand60}.  In much the same way, we can infer that
the eigendistributions of the time evolution in the space of the
Helfand moment have cumulative functions with fractal Hausdorff
dimensions given by a gradient expansion analogous to Eq.
(\ref{Hausdimexp}), where the diffusion coefficient is replaced by
the corresponding transport coefficient. These questions will be
addressed in more detail elsewhere.

\section*{acknowledgments}
The authors wish to acknowledge a very helpful correspondence with
Brian Hunt. T. G. thanks M. Courbage for discussions. P. G. thanks the
National Fund for Scientific Research (FNRS Belgium) and the
InterUniversity Attraction Pole Program of the Belgian Federal
Office of Scientific, Technical and Cultural Affairs for financial
support. JRD thanks the National Science Foundation for support
under grant PHY 98-20824.


\begin{thebibliography}{10}

\bibitem{baranyai93}
Baranyai A., Evans D.~J., and Cohen E.~G.~D., 1993.
\newblock Field dependent conductivity and diffusion in a two-dimensional
  {L}orentz gas.
\newblock {\em Journal of Statistical Physics} {\bf 70}, 1085.

\bibitem{breymann96}
Breymann W., T\'el T., and Vollmer J., 1996.
\newblock Entropy production for open dynamical systems.
\newblock {\em Physical Review Letters} {\bf 77}, 2945.

\bibitem{breymann98}
Breymann W., T\'el T., and Vollmer J., 1998.
\newblock Entropy balance, time reversibility, and mass transport in dynamical
  systems.
\newblock {\em Chaos} {\bf 8}, 396.

\bibitem{chernov93a}
Chernov N.~I., Eyink G.~L., Lebowitz J.~L., and Sinai Ya.~G., 1993.
\newblock Derivation of Ohm's Law in a Deterministic Mechanical Model.
\newblock {\em Physical Review Letters} {\bf 70}, 2209.

\bibitem{chernov93}
Chernov N.~I., Eyink G.~L., Lebowitz J.~L., and Sinai Ya.~G., 1993.
\newblock Steady state electric conductivity in the periodic {L}orentz gas.
\newblock {\em Communications in Mathematical Physics} {\bf 154}, 569.

\bibitem{chernov97a}
Chernov N.~I. and Markarian R., 1997.
\newblock Ergodic properties of Anosov maps with rectangular holes.
\newblock {\em Boletim da Sociedade Brasileira de Matem\'atica} {\bf
28}, 271.

\bibitem{chernov97b}
Chernov N.~I. and Markarian R., 1997.
\newblock Anosov maps with rectangular holes. Nonergodic cases.
\newblock {\em Boletim da Sociedade Brasileira de Matem\'atica} {\bf
28}, 342.

\bibitem{dellago96}
Dellago Ch., Posch H.~A., and Hoover W.~G., 1996.
\newblock Lyapunov instability in a system of hard disks in equilibrium and
nonequilibrium steady states.
\newblock {\em Physical Review E} {\bf 53}, 1485.

\bibitem{dettmann00}
Dettmann C.~P., 2000.
\newblock The Lorentz gas as a paradigm for nonequilibrium stationary states.
\newblock In:  D. Szasz, Editor.
\newblock {\em Hard Ball Systems and Lorentz Gas}.
\newblock Encycl. Math. Sci. (Berlin: Springer).

\bibitem{dorfman95}
Dorfman J.~R. and Gaspard P., 1995.
\newblock Chaotic scattering theory of transport and reaction-rate
  coefficients.
\newblock {\em Physical Review E} {\bf 51}, 28.

\bibitem{dorfman99}
Dorfman J.~R., 1999.
\newblock An Introduction to Chaos in Nonequilibrium Statistical Mechanics.
\newblock (Cambridge UK: Cambridge University Press).

\bibitem{eckmann85}
Eckmann J.-P. and Ruelle D., 1985.
\newblock Ergodic theory of chaos and strange attractors.
\newblock {\em Review of Modern Physics} {\bf 57}, 617.

\bibitem{edgar93}
Edgar G. A., Editor, 1993.
\newblock {\em Classics on Fractals}.
\newblock (Reading MA: Addison-Wesley Publ. Co.).

\bibitem{evans90}
Evans D.~J. and Morriss G.~P., 1990.
\newblock {\em Statistical Mechanics of Nonequilibrium Liquids}.
\newblock (London: Academic Press).

\bibitem{evans90b}
Evans D.~J., Cohen E.~G.~D., and Morriss G.~P., 1990.
\newblock Viscosity of a simple fluid from its maximal {L}yapunov exponents.
\newblock {\em Physical Review A} {\bf 42}, 5990.

\bibitem{evans93}
Evans D.~J., Cohen E.~G.~D., and Morriss G.~P., 1993.
\newblock Probability of second law violations in shearing steady states.
\newblock {\em Physical Review Letters} {\bf 71}, 2401.

\bibitem{gaspard90}
Gaspard P. and Nicolis G., 1990.
\newblock Transport Properties, Lyapunov Exponents, and Entropy per Unit Time.
\newblock {\em Physical Review Letters} {\bf 65}, 1693.

\bibitem{gaspard92}
Gaspard P., 1992.
\newblock Diffusion, effusion, and chaotic scattering: An exactly solvable
  {L}iouvillian dynamics.
\newblock {\em Journal of Statistical Physics} {\bf 68}, 673.

\bibitem{gaspard95c}
Gaspard P. and Baras F., 1995.
\newblock Chaotic scattering and diffusion in the Lorentz gas.
\newblock {\em Physical Review E} {\bf 51}, 5332.

\bibitem{gaspard95}
Gaspard P. and Dorfman J.~R., 1995.
\newblock Chaotic scattering theory, thermodynamic formalism and transport
  coefficients.
\newblock {\em Physical Review E} {\bf 52}, 3525.

\bibitem{gaspard96}
Gaspard P., 1996.
\newblock Hydrodynamic modes as singular eigenstates of the {L}iouvillian
  dynamics: Deterministic diffusion.
\newblock {\em Physical Review E} {\bf 53}, 4379.

\bibitem{gaspard97}
Gaspard P., 1997.
\newblock Entropy production in open, volume preserving systems.
\newblock {\em Journal of Statistical Physics} {\bf 89}, 1215.

\bibitem{gaspard98}
Gaspard P., 1998.
\newblock {\em Chaos, Scattering, and Statistical Mechanics}.
\newblock (Cambridge UK: Cambridge University Press).

\bibitem{gilbert99d}
Gilbert T., 1999.
\newblock {\em Irreversible thermodynamics of reversible dynamical systems}.
\newblock (PhD thesis, University of Maryland, College Park, Maryland, USA).

\bibitem{gilbert99b}
Gilbert T. and Dorfman J.~R., 1999.
\newblock Entropy production ,  From open volume preserving to dissipative
  systems.
\newblock {\em Journal of Statistical Physics} {\bf 96}, 225.

\bibitem{gilbert99}
Gilbert T., Ferguson C.~D., and Dorfman J.~R., 1999.
\newblock Field driven thermostated system~,  a non-linear multi-baker map.
\newblock {\em Physical Review E} {\bf 59}, 364.

\bibitem{gilbert00c}
Gilbert T., Dorfman J.~R., and Gaspard P., 2000.
\newblock Entropy production, fractals, and relaxation to equilibrium.
\newblock submitted to {\em Physical Review Letters}.

\bibitem{helfand60}
Helfand E., 1960.
\newblock Transport Coefficients from Dissipation in a Canonical Ensemble.
\newblock {\em Physical Review} {\bf 119}, 1.

\bibitem{hoover91}
Hoover W.~G., 1991.
\newblock {\em Computational Statistical Mechanics}.
\newblock (Amsterdam: Elsevier Science Publishers).

\bibitem{kantz85}
Kantz H. and Grassberger P., 1985.
\newblock Repellers, semi-attractors, and long-lived chaotic transients.
\newblock {\em Physica D} {\bf 17}, 75.

\bibitem{vonkoch04}
von Koch H., 1904.
\newblock Sur une courbe continue sans tangente obtenue par une construction
  g\'eom\'etrique \'el\'ementaire.
\newblock {\em Arkiv f\"or Mathematik, Astronomi och Fysik} {\bf 1}, 681.
Translated in
\cite{edgar93}.

\bibitem{vonkoch06}
von Koch H., 1906.
\newblock Une m\'ethode g\'eom\'etrique \'el\'ementaire pour l'\'etude de
  certaines questions de la th\'eorie des courbes planes.
\newblock {\em Acta Mathematica} {\bf 30}, 145.

\bibitem{levy38}
L\'evy P., 1938.
\newblock Les courbes planes ou gauches et les surfaces compos\'ees de parties
semblables au tout.
\newblock {\em Journal de l'Ecole Polytechnique}, s\'erie III, {\bf 7-8},
pp. 227--247,
pp. 249--291. Translated in \cite{edgar93}.

\bibitem{pianigiani79}
Pianigiani G. and Yorke J., 1979.
\newblock Expanding maps on sets which are almost invariant,  decay and chaos.
\newblock {\em Trans. Amer. Math. Soc.} {\bf 252}, 351.

\bibitem{posch88}
Posch H.~A. and Hoover W.~G., 1988.
\newblock Lyapunov instability of dense {L}ennard-{J}ones fluids.
\newblock {\em Physical Review A} {\bf 38}, 473.

\bibitem{derham57}
de~Rham G., 1957.
\newblock Sur un exemple de fonction continue sans d\'eriv\'ee.
\newblock {\em Enseignements Math\'ematiques} {\bf 3}, 71. Translated in
\cite{edgar93}.

\bibitem{sinai79}
Sinai Ya.~G., 1979.
\newblock Development of Krylov's idea. Afterwards to,
\newblock N.~S. Krylov, 1979.
\newblock {\em Works on the foundations of statistical physics}.
\newblock (Princeton: Princeton University Press) pp. 239--281.

\bibitem{takagi03}
Takagi T., 1903.
\newblock A simple example of the continuous function without derivative.
\newblock {\em Proceedings of the Physico-Mathematical Society of Japan} {\bf
  1}, 176.

\bibitem{tasaki93}
Tasaki S., Antoniou I., and Suchanecki Z., 1993.
\newblock Deterministic diffusion, de Rham equation and fractal eigenvectors.
\newblock {\em Physics Letters A} {\bf 179}, 97.

\bibitem{tasaki94}
Tasaki S. and Gaspard P., 1994.
\newblock Fractal distribution and {F}ick's law in a reversible chaotic system.
\newblock In: M.~Yamaguti, Editor.
\newblock {\em Towards the Harnessing of Chaos}
(Amsterdam: Elsevier) pp. 273--288.

\bibitem{tasaki95}
Tasaki S. and Gaspard P., 1995.
\newblock Fick's law and fractality of nonequilibrium stationary states in a
  reversible multibaker map.
\newblock {\em Journal of Statistical Physics} {\bf 81}, 935.

\bibitem{tasaki98}
Tasaki S., Gilbert T., and Dorfman J.~R., 1998.
\newblock An analytic construction of the {SRB} measures for baker-type maps.
\newblock {\em Chaos} {\bf 8}, 424.

\bibitem{tasaki99}
Tasaki S. and Gaspard P., 1999.
\newblock Thermodynamic behavior of an area-preserving multibaker map
with energy.
\newblock {\em Theoretical Chemistry Accounts} {\bf 102}, 385.

\bibitem{tel96}
T\'el T., Vollmer J., and Breymann W., 1996.
\newblock {\em Europhysics Letters} {\bf 35}, 659.

\bibitem{tel00}
T\'el T. and Vollmer J., 2000.
\newblock Entropy Balance, Multibaker Maps, and the Dynamics of the Lorentz
Gas.
\newblock In:  D. Szasz, Editor.
\newblock {\em Hard Ball Systems and Lorentz Gas}.
\newblock Encycl. Math. Sci. (Berlin: Springer).

\bibitem{vanbeijeren95}
van Beijeren H. and Dorfman J.~R., 1995.
\newblock Lyapunov exponents and {K}olmogorov-{S}inai entropy for the {L}orentz
  gas at low densities.
\newblock {\em Physical Review Letters} {\bf 74}, 4412.

\bibitem{vanbeijeren96}
van Beijeren H. and Dorfman J.~R., 1996.
\newblock Lyapunov exponents and {K}olmogorov-{S}inai entropy for the {L}orentz
  gas at low densities,  Erratum.
\newblock {\em Physical Review Letters} {\bf 76}, 3238.

\bibitem{vanbeijeren97}
van Beijeren H., Dorfman J.~R., Posch H. A., and Dellago Ch., 1997.
\newblock Kolmogorov-Sinai entropy for dilute gases in equilibrium.
\newblock {\em Physical Review E} {\bf 56}, 5272.

\bibitem{vollmer97}
Vollmer J., T\'el T., and Breymann W., 1997.
\newblock Equivalence of irreversible entropy production in driven systems:  An
  elementary chaotic map approach.
\newblock {\em Physical Review Letters} {\bf 79}, 2759.

\bibitem{vollmer98}
Vollmer J., T\'el T., and Breymann W., 1998.
\newblock Entropy balance in the presence of drift and diffusive currents: An
  elementary map approach.
\newblock {\em Physical Review E} {\bf 58}, 1672.

\end{thebibliography}
\end{document}